\documentclass[manuscript]{acmart}

\AtBeginDocument{%
  }

\setcopyright{acmcopyright}
\copyrightyear{2025}
\acmYear{2025}
\acmDOI{XXXXXXX.XXXXXXX}

\usepackage{booktabs}
\usepackage{multirow}
\usepackage{siunitx}
\usepackage{threeparttable}
\usepackage{makecell}   
\usepackage{graphicx}
\usepackage{caption} 
\usepackage{array}

\begin{document}

\title{Uncovering Students' Inquiry Patterns in GenAI-Supported Clinical Practice: An Integration of Epistemic Network Analysis and Sequential Pattern Mining}
\renewcommand{\shorttitle}{Uncovering Students' Inquiry Patterns in GenAI-Supported Clinical Practice}

\author{Jiameng Wei}
\email{jiameng.wei@monash.edu}
\affiliation{%
  \institution{Monash University}
  \country{Australia}}

\author{Dinh Dang}
\email{ddan0023@student.monash.edu}
\affiliation{%
  \institution{Monash University}
  \country{Australia}}

\author{Kaixun Yang}
\email{kaixun.yang1@monash.edu}
\affiliation{%
  \institution{Monash University}
  \country{Australia}}

\author{Emily Stokes}
\email{emily.stokes@monash.edu}
\affiliation{%
  \institution{Monash University}
  \country{Australia}}

\author{Amna Mazeh}
\email{amna.mazeh@monash.edu}
\affiliation{%
  \institution{Monash University}
  \country{Australia}}

\author{Angelina Lim}
\email{angelina.lim@monash.edu}
\affiliation{%
  \institution{Monash University}
  \country{Australia}}

\author{David Wei Dai}
\email{david.dai@ucl.ac.uk}
\affiliation{%
  \institution{University College London}
  \country{United Kingdom}}

\author{Joel Moore}
\email{Joel.Moore@monash.edu}
\affiliation{%
  \institution{Monash University}
  \country{Australia}}

\author{Yizhou Fan}
\email{fyz@pku.edu.cn}
\affiliation{%
  \institution{Peking University}
  \country{China}}

\author{Danijela Gasevic}
\email{Danijela.Gasevic@monash.edu}
\affiliation{%
  \institution{Monash University}
  \country{Australia}}

\author{Dragan Gasevic}
\email{dragan.gasevic@monash.edu}
\affiliation{%
  \institution{Monash University}
  \country{Australia}}

\author{Guanliang Chen}
\authornote{Corresponding author.}
\email{guanliang.chen@monash.edu}
\affiliation{%
  \institution{Monash University}
  \country{Australia}}

\renewcommand{\shortauthors}{Wei et al.}

\begin{abstract}
Assessment of medication history-taking has traditionally relied on human observation, limiting scalability and detailed performance data. While Generative AI (GenAI) platforms enable extensive data collection and learning analytics provide powerful methods for analyzing educational traces, these approaches remain largely underexplored in pharmacy clinical training. This study addresses this gap by applying learning analytics to understand how students develop clinical communication competencies with GenAI-powered virtual patients---a crucial endeavor given the diversity of student cohorts, varying language backgrounds, and the limited opportunities for individualized feedback in traditional training settings. We analyzed 323 students' interaction logs across Australian and Malaysian institutions, comprising 50,871 coded utterances from 1,487 student-GenAI dialogues. Combining Epistemic Network Analysis to model inquiry co-occurrences with Sequential Pattern Mining to capture temporal sequences, we found that high performers demonstrated strategic deployment of information recognition behaviors. Specifically, high performers centered inquiry on recognizing clinically relevant information, integrating rapport-building and structural organization, while low performers remained in routine question-verification loops. Demographic factors including first-language background, prior pharmacy work experience, and institutional context, also shaped distinct inquiry patterns. These findings reveal inquiry patterns that may indicate clinical reasoning development in GenAI-assisted contexts, providing methodological insights for health professions education assessment and informing adaptive GenAI system design that supports diverse learning pathways.
\end{abstract}





\maketitle

\section{Introduction}
Medication history-taking represents a fundamental competency in health professions education, encompassing the systematic processes through which clinicians identify medication needs, review current therapies, and synthesize information to ensure safe prescribing practices \citep{bajis2019pharmacy}. Within pharmacy education, medication history-taking competencies can be understood as a task-specific subset of broader clinical communication skills. The Calgary-Cambridge model provides a comprehensive framework for clinical communication skills, including rapport-building, strategic questioning techniques, and recognition of verbal and non-verbal cues \citep{silverman2016skills}, within which medication history-taking requires practitioners to demonstrate clinical reasoning through structured inquiry and adaptive responses based on patient cues \citep{keifenheim2015teaching, higgs2024clinical}. While these integrated communication and reasoning skills require extensive structured training to develop to professional standards \citep{gortney2019clinical, francis2023accuracy}, traditional supervised practices restrict students to a few patient encounters through role-play activities, with performance evaluation heavily relying on manual observation that provides limited behavioral data for skill assessment \citep{peart2022clinical, pottle2019virtual}.

Generative AI (GenAI) holds the promise to transform clinical communication training by enabling scalable and dynamic practice environments through AI-powered virtual patients (VPs) \citep{zamanifar2025application}. Such GenAI systems demonstrate natural language capabilities that create fluid and contextually appropriate patient interactions \citep{rabbani2025generative}. These platforms generate personalized learning experiences with multimodal capabilities and are now leveraged for pharmacy training, producing extensive interaction logs that capture detailed student-VP dialogues \citep{mortlock2024generative, zawiah2023chatgpt, stokes2024harnessing}. Understanding how students develop clinical competencies through these AI interactions presents both an opportunity and a methodological challenge. While GenAI platforms capture thousands of interaction traces revealing moment-by-moment learning processes, traditional assessment methods (interviews, questionnaires, and direct observation) cannot systematically analyze these temporal patterns at scale \citep{jovanovic2020chatbots}. This analytical gap necessitates the use of learning analytics approaches to transform raw interaction data into meaningful insights about clinical reasoning development for supporting effective teaching and learning practices \citep{lee2020effective}.

Learning analytics has been increasingly applied in health professions education to examine various clinical competencies. Studies using Epistemic Network Analysis (ENA) have revealed how nursing students develop safety competencies and how procedural learning varies by experience levels \citep{shah2021modeling, ruis2019multiple}. Transitional Network Analysis has captured temporal dynamics in students' interactions with VPs, identifying recurring behavioral patterns that shape learning outcomes \citep{mihtsun2025analysis}. These studies demonstrate how learning analytics can uncover relationships between behavioral patterns, learner characteristics, and performance outcomes in traditional or scripted simulation environments. However, existing studies primarily rely on single analytical approaches that capture either co-occurrence patterns or temporal transitions of students' behaviors in isolation, limiting their ability to reveal both the structural relationships and sequential progressions that characterize clinical reasoning development. Moreover, while these methods have proven valuable in traditional training contexts, the emergence of GenAI-powered VPs presents a fundamentally different learning environment where students can engage in natural language dialogues rather than following pre-scripted scenarios. Research has established that clinical reasoning patterns vary by performance level \citep{furstenberg2020assessing} and learner characteristics affect communication outcomes in traditional settings \citep{groene2022attitude}, yet no studies have applied learning analytics to examine these patterns in GenAI-supported medication history-taking. This represents a critical gap given the accelerating adoption of GenAI in health professions education, as understanding these patterns through combined analytical approaches is essential for optimizing AI-based training systems.

To address these gaps, this study aimed to examine how pharmacy students develop medication history-taking competencies through GenAI-supported practices. Formally, this study was guided by the \textbf{Research Question}: \textit{What inquiry patterns emerge from students' interactions with GenAI-powered VPs, and how do these patterns relate to students' performance in clinical examinations and their demographic characteristics?} To answer this research question, we analyzed 323 students' interaction logs from a GenAI-powered platform where students practice patient interviews with VPs. The 50,871 utterances from 1,487 student-VP dialogues were coded based on a carefully-crafted codebook consisting of seven codes capturing social, procedural, and clinical dimensions of medication history-taking. Using these coded data, we applied two complementary methods: ENA to examine how different inquiry types co-occur in pairs, and Sequential Pattern Mining (SPM) to reveal longer sequences of multiple inquiry behaviors over time, together capturing how students' inquiry strategies evolve across different performance levels and demographic groups during GenAI-simulated patient interactions. The findings inform in-depth insights for designing clinical GenAI-powered training systems and demonstrate how learning analytics can evaluate GenAI-supported health professions education.

\section{Background}
\subsection{Medication History-Taking: From Traditional Training to GenAI-Enabled Practice}
Medication history-taking through structured patient or carer interviews and verification from multiple sources (e.g., General Practitioner records, community pharmacy dispensing histories) represents a core skill in pharmacist training, requiring comprehensive medication information gathering while identifying and documenting discrepancies \citep{stark2020implementing, bajis2019pharmacy}. The Calgary-Cambridge framework guides effective practices through strategic transitions between standardized questioning protocols and adaptive follow-up techniques \citep{silverman2016skills}. Despite this established framework, pharmacy students often miss medications or record incorrect dosages, with studies demonstrating continued need for pharmacist verification to ensure patient safety \citep{francis2023accuracy}. Traditional training through supervised clinical encounters faces significant logistical constraints: high costs and organizational complexity limit students to single patient interactions, preventing essential repetitive practice \citep{dai2025generative, pottle2019virtual}. Additionally, supervised environments can inadvertently impact less confident students' performance, while subjective observational assessment fails to capture individual development trajectories \citep{paradis2017beyond}. The growing adoption of simulation-based training in pharmacy education has partially addressed these limitations, enabling scalable, reproducible learning environments with automated tracking and consistent feedback mechanisms \citep{mcbane2023overview, ventola2019virtual, elendu2024impact}. 

Traditional simulation platforms with pre-scripted VPs, though an improvement, remain constrained by rigid dialogue structures and limited response variability. The emergence of GenAI-powered VPs holds promise to fundamentally transform training possibilities by generating contextually adaptive, authentic dialogue responses that more closely approximate real patient interactions \citep{moser2025twelve}. These GenAI systems have demonstrated tangible benefits in clinical training: enabling natural language dialogue for history-taking practices \citep{maicher2023artificial, holderried2024generative}, providing real-time feedback that improves the performance of Objective Structured Clinical Examinations (OSCEs) \citep{stokes2024harnessing}, enhancing students' confidence in motivational interviewing \citep{sasser2025ai}, and supporting development of patient counseling skills \citep{nakagawa2022communication}. Moreover, these GenAI-powered systems generate extensive interaction logs capturing complete student-VP dialogues, creating unprecedented opportunities for understanding students' learning processes in this particular setting \citep{ali2025will}. Yet, these studies primarily focus on system development and measuring performance improvements in clinical assessments rather than analyzing students' behavioral patterns and needs during the learning processes.

\subsection{Learning Analytics in Health Professions Education}
Learning analytics methods have proven valuable for analyzing various aspects of clinical training in health professions education. Early work in diagnostic reasoning with VPs demonstrated how student performance can be decomposed into dimensions such as information gathering, hypothesis generation, and decision-making accuracy \citep{furlan2022learning}. In nursing team simulations, ENA has been employed to encode spatial and behavioral traces, producing interpretable visualizations that teachers evaluated as useful for reflecting on collaboration and performance \citep{fernandez2021modelling}. More recently, multimodal dashboards have been developed to support teacher-guided reflection in clinical simulations, demonstrating how learning analytics can connect data-informed insights with authentic debriefing practices in healthcare training contexts \citep{echeverria2024teamslides}. Complementary studies have shown that team communication and performance in healthcare simulations can be modeled from multimodal signals, highlighting the feasibility of generating near-real-time indicators of clinical processes \citep{zhao2022modelling}. However, despite these advances, existing research has not yet leveraged the detailed utterance-level and temporal traces collected through interaction with GenAI-powered VP platforms to examine how students develop clinical communication competencies. Recent work has demonstrated that ENA and sequence-based analyses offer complementary perspectives on learner behavior, with ENA revealing co-occurrence structures among coded elements and SPM detecting extended temporal sequences \citep{prakash2025decoding}. This combined analytical approach is particularly well-suited for analyzing the rich conversational data generated in GenAI-mediated clinical training, where students' inquiry patterns unfold through unrestricted natural language interactions.

\subsection{Clinical Inquiry Patterns in History-Taking}
To investigate students' inquiry behaviors in this new GenAI-assisted setting, we developed hypotheses 
by linking findings from existing research on clinical communication education to the unique affordances of GenAI environments. Prior to GenAI's emergence, medication history-taking research often examined students' inquiry patterns from four key aspects: student performance levels \citep{furstenberg2020assessing}, first-language background \citep{venkat2020using}, prior pharmacy work experience \citep{choi2023prior}, and institutional context \citep{jessee2016influences}. GenAI's unique capabilities, particularly enabling students to engage in varied and nuanced natural language dialogues, create a novel context requiring systematic investigation \citep{rabbani2025generative}. Building on the Calgary-Cambridge framework and clinical reasoning indicators \citep{silverman2016skills, furstenberg2020assessing}, we formulate the following hypotheses examining how these factors manifest in students' interactions with GenAI-powered VPs:

\subsubsection{Performance Differences in Clinical Reasoning}
Student performance in medication history-taking reflects different levels of clinical reasoning maturity, characterized by systematic progressions and cognitive organization that enable practitioners to adapt approaches based on contextual demands \citep{higgs2024clinical, connor2020clinical}. This maturity in clinical reasoning translates to employing diverse questions within logical sequences, from open-ended exploration to targeted verification \citep{peart2022clinical}. \citet{furstenberg2020assessing} validated empirical indicators for clinical reasoning assessment, identifying logical question ordering, systematic information gathering, and pattern recognition as competence markers. We therefore hypothesize: \underline{\textbf{H1.1}} \textit{compared to high performers, students with lower performance confine their inquiry patterns to relatively fewer inquiry types}. Additionally, information recognition represents a fundamental expertise component, enabling clinicians to identify and prioritize relevant information from patient narratives and respond appropriately \citep{pelaccia2019simulation}, with research showing that the frequency of recognizing relevant cues differentiates expert from novice practitioners \citep{peart2022clinical}. In pharmacy education, identifying medication-related problems and interpreting patient information constitute essential clinical reasoning indicators \citep{altalhi2021development}. We therefore hypothesize: \underline{\textbf{H1.2}} \textit{compared to low performers, high performers adopt information recognition more frequently as a central element in their inquiries}. Recent analysis of effective digital VP interactions revealed systematic transitions from social exchanges (e.g., greeting, establishing rapport) to focused clinical discussions (e.g., exploring symptoms and medication history) \citep{mihtsun2025analysis}. We therefore hypothesize: \underline{\textbf{H1.3}} \textit{compared to low performers, high performers demonstrate more transitions from social exchanges to focused clinical discussions}. 

\subsubsection{First-Language Background} 
English-as-an-Additional-Language (EAL) learners face increased cognitive demands processing clinical information in their second language, demonstrating lower communication scores in OSCEs \citep{venkat2020using}. This cognitive load constrains their ability to simultaneously manage linguistic processing and complex clinical reasoning, often resulting in more cautious and verification-focused communication strategies \citep{ziaei2015internationally}. We therefore hypothesize: \underline{\textbf{H2.1}} \textit{compared to First language (L1) speakers of English, EAL learners demonstrate more verification-focused inquiries}. L1 speakers, conversely, adopt more patient-oriented elements such as rapport-building, collaborative dialogue, and responsive engagement in their clinical interactions \citep{groene2022attitude}. We therefore hypothesize: \underline{\textbf{H2.2}} \textit{compared to EAL learners, L1 speakers demonstrate more patient-oriented inquiries}.

\subsubsection{Prior Pharmacy Work Experience} 
Students with prior pharmacy work experience demonstrate enhanced communication performance in their educational training \citep{choi2023prior}, and those who worked part-time during undergraduate studies show higher empathy levels \citep{halimi2023resilience}. We therefore hypothesize: \underline{\textbf{H3.1}} \textit{compared to students without prior pharmacy work experience, students with prior experience incorporate more social elements in their clinical inquiries}. Moreover, clinical experience develops metacognitive capabilities through reflective inquiry (e.g., synthesis and evaluation of information collected) and self-regulation \citep{higgs2024clinical}. In contrast, students without prior work experience often demonstrate poorer knowledge retention in clinical interactions, struggling to store and retrieve important information effectively \citep{valdez2013impact}. We therefore hypothesize: \underline{\textbf{H3.2}} \textit{compared to students without pharmacy work experience, students with prior experience demonstrate enhanced metacognitive capabilities through more information synthesis patterns.}

\subsubsection{Institutional Context}
Research shows that local institutional factors, including student demographics, faculty teaching styles, and campus-specific learning cultures, significantly influence how students develop clinical communication skills \citep{jessee2016influences}. Healthcare communication training varies across institutional contexts, with evidence from the same university showing significant differences between different campuses (e.g., Australian and Malaysian) even when implementing identical educational activities \citep{brock2020implementing}. Whether these contextual differences persist when using identical GenAI-powered training platforms remains to be investigated. We therefore hypothesize: \underline{\textbf{H4}} \textit{students within different institutional contexts produce distinct inquiry patterns}.

\section{Method}
\subsection{Study Context}
This study was conducted during Semester 1 (February-May 2024) within a second-year professional practice course of an undergraduate pharmacy program at a large Australian university, delivered across two campuses, Australia and Malaysia (Ethics approval: Project ID 41702). The course builds on students’ understanding of the pharmacist’s role within the healthcare system and introduces key components of a standardized approach to patient-centered care. Core to the course is developing competencies in obtaining and documenting a Best Possible Medication History (BPMH), which is assessed through OSCE stations where students demonstrate their ability to gather and verify complete medication profiles, alongside clinical reasoning and safe medicine supply procedures \citep{lim2024using}. 

\begin{figure} [h!]
\centering 
\includegraphics[width=0.7\linewidth]{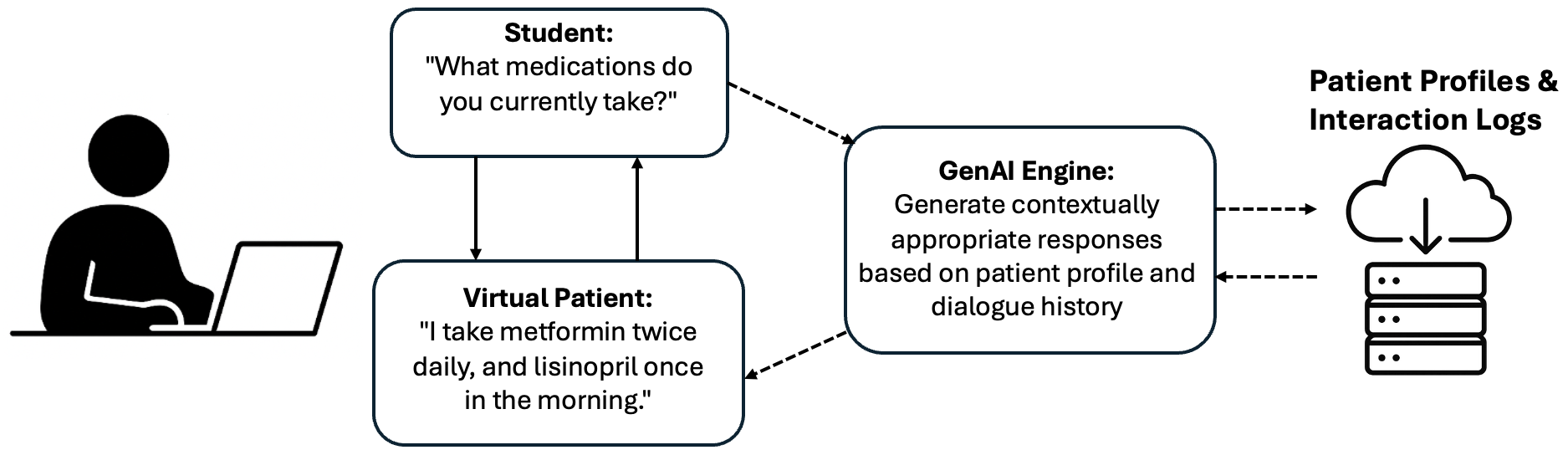} 
\caption{Clinical simulation platform architecture illustrating the student-VP interaction cycle, with GenAI engine processing and data logging components (solid arrows: user interactions; dashed arrows: system processing).} 
\label{fig:inter}
\end{figure}

To support students in practicing BPMH competencies, a GenAI-powered simulation platform was offered as a supplementary tool for students to access at their preferred time and location to strengthen core skills required for future clinical practices. The platform, powered by GPT-4, enabled voice-based interactions where students verbally questioned AI avatars configured as patients with specific medication profiles, receiving real-time spoken responses. All verbal interactions were automatically transcribed by the platform into textual format for subsequent analysis in this study (Figure \ref{fig:inter}).   

\subsection{Dataset}\label{sec:subsection:dataset}
This study utilized a multi-source dataset comprising students’ interaction logs with the GenAI-powered training platform described above, demographic information, and performance outcomes. Specifically, a total of 323 pharmacy students (Australia: n=211; Malaysia: n=112) engaged with the platform, generating 1,487 dialogue sessions. Each student-VP dialogue generated detailed log data, with all verbal communications transcribed into textual utterances for analysis. Together, these sessions contained 50,871 utterances capturing complete student-VP exchanges (median=35 utterances per session, IQR=22.5). To address hypotheses H1.1-H1.3, we collected students’ oral performance scores from the BPMH station, operationalized through the OSCEs. These scores, ranging from 1 to 100, were assigned by two independent assessors and served as a measure of students’ communication competence in medication history-taking. To address hypotheses H2.1-H2.2, H3.1-H3.2, and H4, we retrieved students’ demographic data collected through a pre-study survey, capturing first-language background (35.9\% L1 speakers; 64.1\% EAL learners), prior pharmacy work experience (54.8\% with experience; 45.2\% without experience), and institutional context (65.3\% Australia vs. 34.7\% Malaysia).

To analyze the 50,871 utterances and identify inquiry patterns, we developed a coding scheme by integrating two complementary frameworks. The Calgary-Cambridge consultation model \citep{silverman2016skills} provided comprehensive, high-level guidance for clinical communication skill development and assessment, offering broad categories such as ``gathering information'' and ``building relationship.'' However, these high-level categories alone could not capture the specific inquiry intentions within individual utterances. Therefore, we complemented it with the Professional Communication Rubric \citep{barnett2020expert}, which offers more granular behavioral descriptors that could be operationalized at the utterance level. By combining these frameworks, we derived codes such as ``Statement'' (ST) from ``managing the encounter'' and ``concluding the encounter'' items to capture structural organization utterances, and ``Specifying Symptoms'' (SS) from ``exploring patient's symptoms'' techniques to identify focused medication inquiries. Through iterative analysis of a random sample of 20 student-VP dialogues, we initially identified 13 distinct inquiry types that were subsequently consolidated to seven codes, with complete descriptions and examples provided in Table \ref{tab:inquiry_codes}.

\begin{table}[hbt!]
\caption{Inquiry codes and descriptions with examples}
\label{tab:inquiry_codes}
\resizebox{1\textwidth}{!}{
\begin{tabular}{@{}lll@{}}
\toprule
\textbf{Inquiry Code} & \textbf{Description} & \textbf{Example} \\ \midrule
Routine Question (\textbf{RQ}) & \begin{tabular}[c]{@{}l@{}}Standardized inquiries typically asked of all patients\\ during medication history collection\end{tabular} & \textit{"Do you take any vitamins?"} \\ \midrule
Specifying Symptoms (\textbf{SS}) & \begin{tabular}[c]{@{}l@{}}Focused questions about particular medications,  symptoms, \\ or conditions requiring detailed exploration\end{tabular} & \textit{\begin{tabular}[c]{@{}l@{}}"Are you still taking the Atorvastatin \\ 20mg at night for cholesterol?"\end{tabular}} \\ \midrule
Checking with Patient (\textbf{CP}) & \begin{tabular}[c]{@{}l@{}}Verification of specific medication parameters including \\ dosage, frequency, strength, etc.\end{tabular} & \textit{\begin{tabular}[c]{@{}l@{}}"Could you tell me the strength of \\ this medication?"\end{tabular}} \\ \midrule
\begin{tabular}[c]{@{}l@{}}Recognising and Responding\\ to Relevant Information (\textbf{RRRI})\end{tabular} & \begin{tabular}[c]{@{}l@{}}Identification of clinically significant information and \\ formulation of targeted follow-up inquiries\end{tabular} & \textit{\begin{tabular}[c]{@{}l@{}}"That's correct, 20mg at night. How long \\ have you been taking the Atorvastatin?"\end{tabular}} \\ \midrule
Professional Instructions (\textbf{PI}) & \begin{tabular}[c]{@{}l@{}}Instructions regarding professional follow-up actions \\ based on identified medication issues\end{tabular} & \textit{\begin{tabular}[c]{@{}l@{}}"I'll be sure to talk to your doctor \\ about that."\end{tabular}} \\ \midrule
Statement (\textbf{ST}) & \begin{tabular}[c]{@{}l@{}}Summaries or restatements of patient information without \\ seeking confirmation\end{tabular} & \textit{\begin{tabular}[c]{@{}l@{}}"So you stopped taking the Amlodipine \\ and now take Prazosin 1mg at night."\end{tabular}} \\ \midrule
Chitchat (\textbf{CC}) & \begin{tabular}[c]{@{}l@{}}Social exchanges including greetings, expressions of \\ empathy, and rapport-building communication\end{tabular} & \textit{"Nothing serious, no need to worry."} \\ \bottomrule
\end{tabular}}
\end{table}

Given the substantial size of the dataset, manually coding all 50,871 utterances would have been prohibitively costly and time-consuming. We therefore adopted a two-stage approach: first manually coding a representative subset of data, and then training a machine learning model to automatically infer labels for the remaining unlabeled utterances. Guided by the codebook described above, two annotators with medication history-taking expertise independently coded student utterances through four iterative refinement rounds, comparing labels and resolving discrepancies after each round. Inter-rater reliability progressively improved, with Cohen’s kappa reaching 0.880 in the final round, indicating very good agreement \citep{dettori2020kappa}. After satisfactory reliability was achieved, the annotators labeled an additional set of dialogues, resulting in a total of 4,985 manually coded utterances distributed across the seven inquiry codes (RQ: 31.21\%; CP: 21.02\%; CC: 16.91\%; SS: 13.30\%; ST: 8.59\%; PI: 4.83\%; RRRI: 4.13\%).

Using the manually-labeled set of data, we trained an automatic labeling model by partitioning the data into training (80\%) and testing (20\%) subsets. To enhance classification accuracy, each target utterance was augmented with its immediately preceding and following utterances, thereby incorporating conversational context. We fine-tuned the LLaMA-3 8B language model \citep{dubey2024llama}, selecting the final model based on its performance on the held-out test set. The model achieved an overall accuracy of 0.800 and a weighted F1-score of 0.805. Prediction accuracies by category were: PI=0.915, CC=0.905, RQ=0.894, CP=0.743, SS=0.712, ST=0.628, and RRRI=0.512. While performance was relatively lower for minority codes (ST and RRRI), the model achieved strong performance on the remaining codes, providing sufficiently reliable predictions for large-scale coding. This trained model was then applied to the unlabeled utterances, resulting in a fully coded dataset of 50,871 utterances for subsequent analysis. The occurrence frequency of different inquiry codes is detailed in Table \ref{tab:code_dist_overview}.

\subsection{Data Analysis}
To investigate hypotheses H1.1-H1.3, students were first categorized as \textbf{high performers} and \textbf{low performers} based on their oral BPMH scores. Following the mean-split approach commonly used in prior research \citep{wu2019analysing}, students whose BPMH scores were above the mean (i.e., 90.6 in our case) were classified as high performers, and those below the mean were classified as low performers. For hypotheses H2.1-H2.2, H3.1-H3.2, and H4, students were grouped according to their demographic attributes: \textbf{first-language background} (L1 speakers vs. EAL learners for H2.1-H2.2), \textbf{prior work experience} (students with prior pharmacy work experience vs. those without for H3.1-H3.2), and \textbf{institutional context} (Australian campus vs. Malaysian campus for H4).  

Across all hypotheses, we employed three complementary analytical approaches to examine students’ inquiry behaviors. First, we compared the \textbf{overall occurrence frequencies} of each inquiry code between groups using chi-squared tests with Benjamini-Hochberg correction for multiple comparisons, identifying which inquiry types were over- or under-represented in each group. Second, similar to prior research on health professions education \citep{shah2021modeling, ruis2019multiple}, we applied \textbf{Epistemic Network Analysis (ENA)} to model pairwise co-occurrences of inquiry codes, providing insight into which codes tend to co-appear in close conversational proximity and how their structural relationships differ across groups \citep{woollaston2025archie, wang2023matters}. However, ENA is limited to dyadic associations and does not capture longer progressions of inquiry behaviors \citep{prakash2025decoding}. For instance, if a student were to progress systematically from social rapport through routine questions to detailed medication verification in a CC-RQ-SS sequence, ENA would decompose this into separate CC-RQ, RQ-SS, and CC-SS connections, losing the temporal ordering crucial for understanding strategic inquiry development. Therefore, we employed \textbf{Sequential Pattern Mining (SPM)} \citep{lin2022good} to uncover longer, temporally ordered sequences of multiple inquiry codes that characterize strategic inquiry development within a student-VP dialogue. Combining these three methods enabled a comprehensive view of students’ inquiry patterns: frequency analysis captured broad differences in code distributions, ENA revealed structural co-occurrence networks, and SPM exposed longer sequences that reflected students’ strategic progression through the combinational use of different types of inquiries. Together, these analyses provided a richer and more nuanced understanding of how performance level, language background, prior work experience, and institutional context shape students’ inquiry behaviors during GenAI-powered medication history-taking practices.

\subsubsection{Epistemic Network Analysis}
ENA was used to model the co-occurrence structure of inquiry codes within dialogues, allowing us to identify which codes were most strongly connected across groups. The analysis was performed using the ENA Web Tool (version 1.7.0) \citep{marquart2018epistemic}. Networks were constructed at the dialogue level ($n = 1,487$) and then aggregated by group membership for comparison. We applied a moving stanza window of six utterances to capture local co-occurrences, defining a connection whenever two codes appeared within the same window \citep{siebert2017search}. This window size was determined based on preliminary analysis of the manually-coded data described in Section \ref{sec:subsection:dataset} to balance sensitivity to immediate adjacency with the ability to capture short contextual spans. Differential connection strengths between groups were visualized using subtraction plots, and statistical significance of group differences was assessed by comparing the positions of analysis units (i.e., dialogues) in the reduced ENA space via two-tailed Mann-Whitney U tests, with rank-biserial correlation ($r$) reported as the effect size.

\subsubsection{Sequential Pattern Mining}
To examine extended inquiry trajectories beyond pairwise co-occurrences, each student-GenAI dialogue was transformed into a sequence preserving the temporal order of the corresponding inquiry codes. Frequent sequences were extracted with the \texttt{seqefsub()} function in the \texttt{TraMineR} package in R, using a minimum support threshold of $p_\text{min} = 0.05$ (i.e., appearing in at least 5\% of the dialogues) and retaining patterns with empirical stability (support\textsubscript{p95}  $\geq$ 0.5). Group comparisons were then conducted using the \texttt{seqecmpgroup()} function, which applies Pearson’s chi-squared test to compare the presence of each sequence between groups \citep{lin2022good}. To adjust for multiple comparisons arising from testing many candidate sequences, we employed the Benjamini-Hochberg procedure, which controls the false discovery rate. For each significant sequence, standardized residuals were inspected to determine the group in which it was overrepresented, while effect sizes were quantified using Cramér’s $V$ and complemented with group-wise presence rates to aid interpretation \citep{munk2017quantitative}. 

\begin{table}[!htb]
\caption{Percentages of inquiry codes across student groups. AU/MY denotes the Australia/Malaysia campus. Significant differences ($p<0.05$) are marked using asterisks; bold values indicate the overrepresented groups (higher proportions).}
\label{tab:code_dist_overview}
\resizebox{1\textwidth}{!}{
\begin{tabular}{@{}l c|cc|cc|cc|cc@{}}
\toprule
\multirow{2}{*}{\textbf{Inquiry Code}} & \multirow{2}{*}{\textbf{All}} 
& \multicolumn{2}{c|}{\textbf{Performance}} 
& \multicolumn{2}{c|}{\textbf{First-Language Background}} 
& \multicolumn{2}{c|}{\textbf{Prior Work Experience}} 
& \multicolumn{2}{c}{\textbf{Institutional Context}} \\  
\cmidrule(lr){3-4}\cmidrule(lr){5-6}\cmidrule(lr){7-8}\cmidrule(lr){9-10} &  & High & Low & L1 & EAL & With & Without & AU & MY \\ 
\midrule
Routine Question (RQ) & 37.3 & *36.9 & \textbf{*38.2} & *35.5 & \textbf{*38.2} & *36.9 & \textbf{*37.8} & *36.1 & \textbf{*39.0} \\
Specifying Symptoms (SS) & 13.6 & 13.4 & 14.0 & 13.3 & 13.7 & *12.8 & \textbf{*14.5} & *13.3 & \textbf{*14.0} \\
Checking with Patient (CP) & 20.3 & *19.9 & \textbf{*21.1} & 20.5 & 20.2 & 20.2 & 20.3 & \textbf{*20.8} & *19.6 \\
Recognizing \& Responding to \\ Relevant Information (RRRI) & 3.2 & \textbf{*3.5} & *2.6 & \textbf{*3.5} & *3.0 & \textbf{*3.4} & *3.0 & \textbf{*3.4} & *2.9 \\
Professional Instructions (PI) & 2.2 & 2.3 & 2.1 & 2.4 & 2.2 & 2.3 & 2.1 & *2.1 & \textbf{*2.4} \\
Statement (ST) & 8.1 & \textbf{*8.5} & *7.3 & \textbf{*8.7} & *7.9 & \textbf{*8.5} & *7.7 & \textbf{*8.6} & *7.5 \\
Chitchat (CC) & 15.3 & \textbf{*15.5} & *14.7 & \textbf{*16.2} & *14.8 & \textbf{*15.9} & *14.6 & \textbf{*15.7} & *14.6 \\ \bottomrule
\end{tabular}}
\end{table}

\begin{figure}
  \centering
  \tabcolsep=0pt
  \begin{tabular*}{\textwidth}{@{\extracolsep{\fill}}cccc}
    \includegraphics[width=0.24\textwidth]{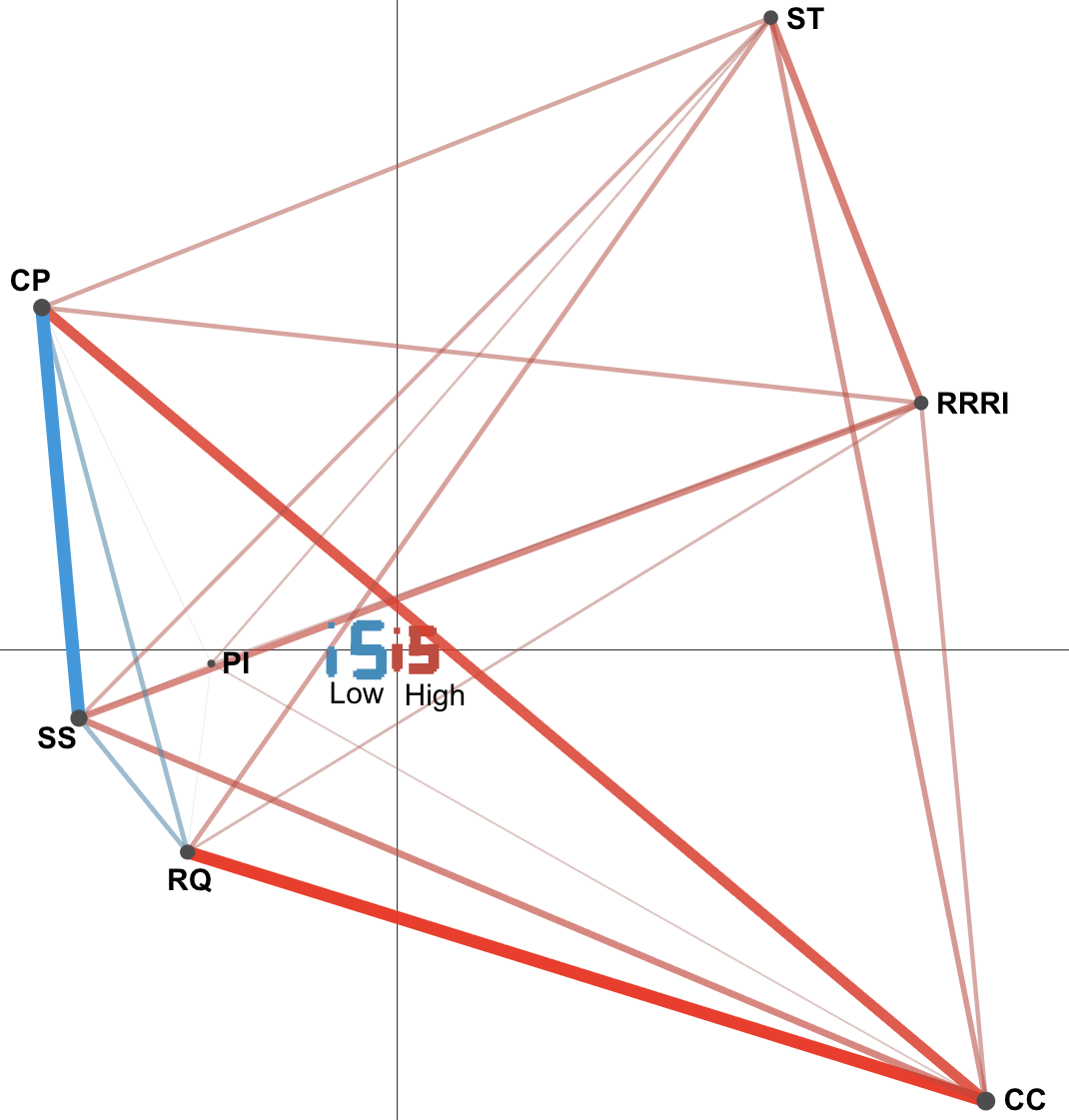} &
    \includegraphics[width=0.24\linewidth]{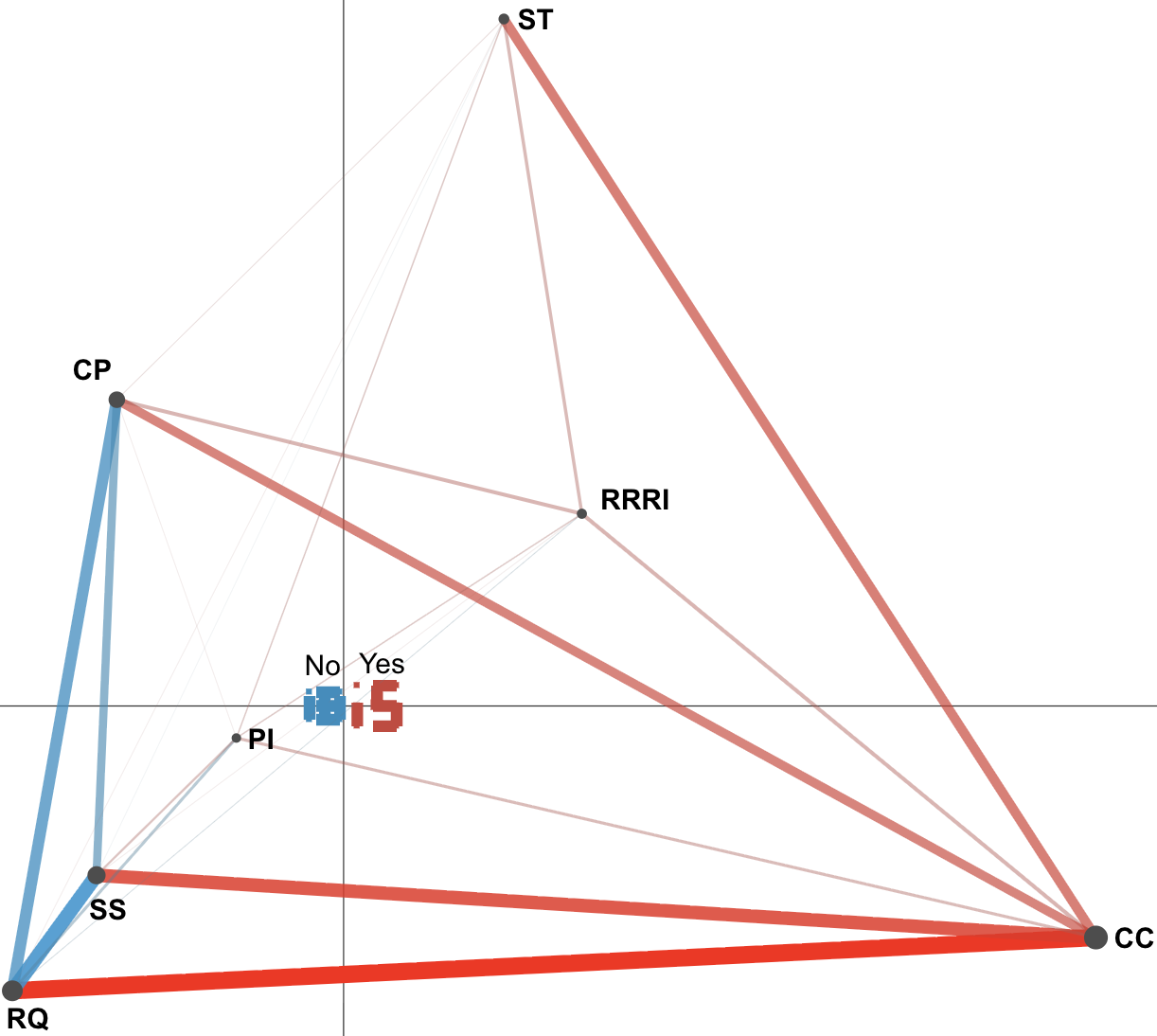} &
    \includegraphics[width=0.24\linewidth]{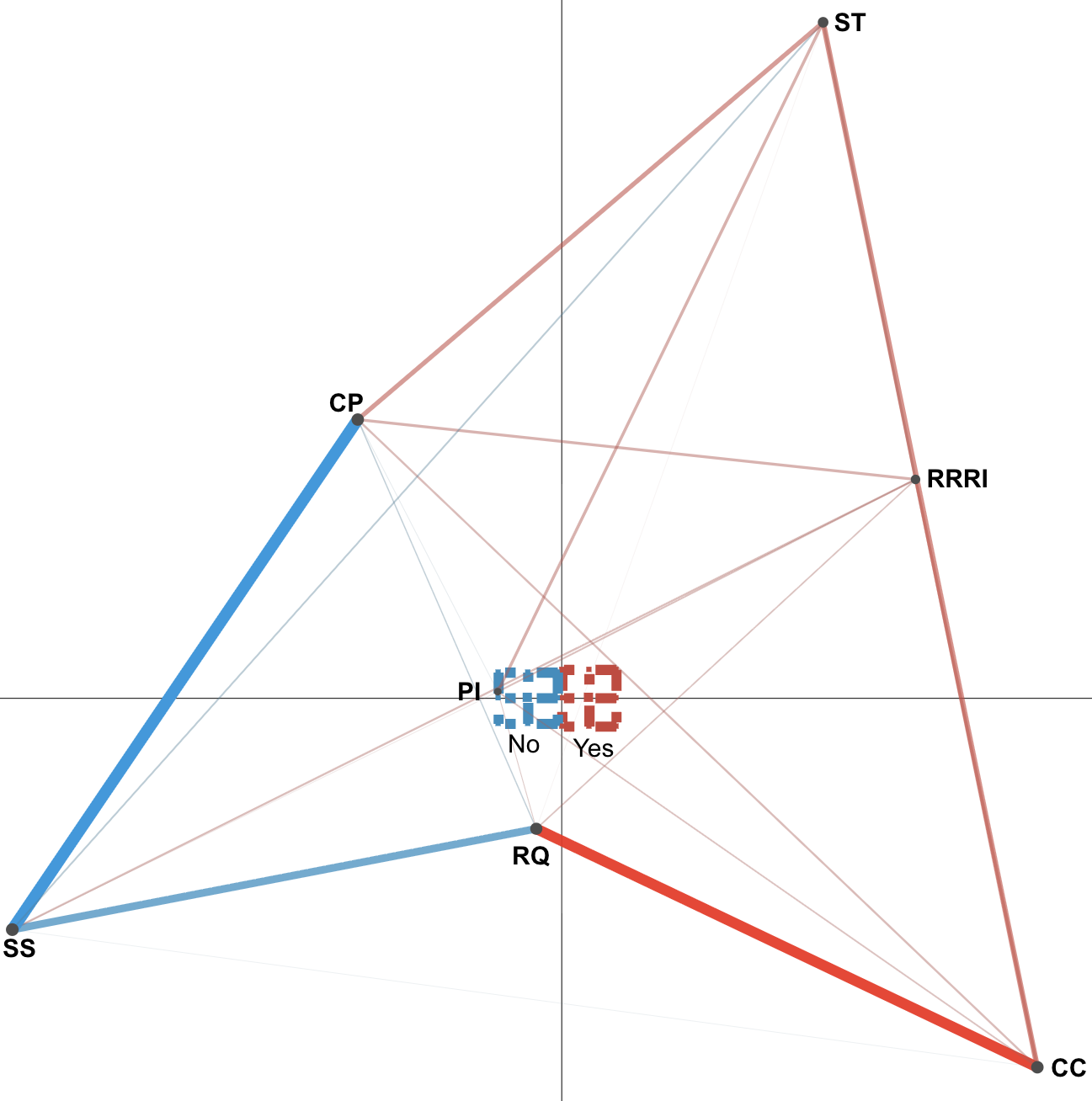} &
    \includegraphics[width=0.24\linewidth]{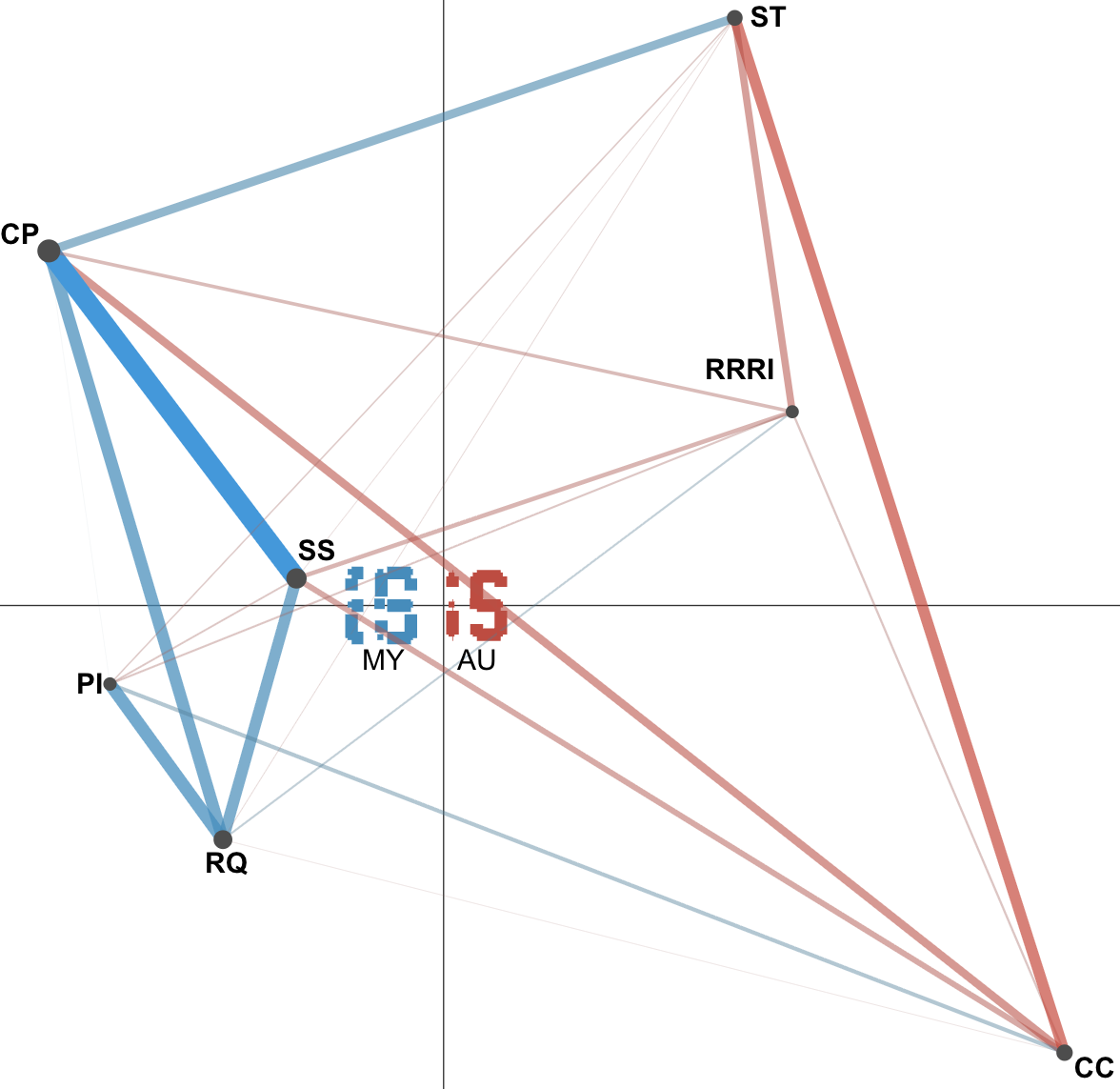} \\   \parbox{0.24\textwidth}{\centering \small (a) Performance\\(Red: High; Blue: Low)} &    \parbox{0.24\textwidth}{\centering \small (b) First-Language Background\\(Red: L1; Blue: EAL)} &    \parbox{0.24\textwidth}{\centering \small (c) Prior Work Experience\\(Red: With; Blue: Without)} &
    \parbox{0.24\textwidth}{\centering \small (d) Institutional Context\\(Red: Australia; Blue: Malaysia)}
  \end{tabular*}
  \caption{Epistemic Network Analysis subtraction plots.}
  \label{fig:ena}
\end{figure}

\begin{table*}[htbp]
\centering
\scriptsize
\setlength{\tabcolsep}{3pt}
\begin{threeparttable}
\caption{Results on Sequential Pattern Mining. To facilitate interpretation of between-group differences, we report only the top ten frequent sequences with the largest effect sizes for each comparison. Percentages denote presence rates in each group. All entries are significant at $p<0.05$ (BH-adjusted). V = Cramér’s V. Overrep = group with higher presence.}
\label{tab:h1h4h5_sm}
\begin{tabular}{@{}p{0.15\linewidth} c c c c | p{0.15\linewidth} c c c c | p{0.15\linewidth} c c c c@{}}
\toprule
\multicolumn{5}{c|}{\textbf{H1.1-H1.3: Performance (High vs.\ Low)}} &
\multicolumn{5}{c|}{\textbf{H3.1-H3.2: Prior Work Experience (With vs.\ Without)}} &
\multicolumn{5}{c}{\textbf{H4: Institutional Context (AU vs.\ MY)}} \\
\cmidrule(lr){1-5} \cmidrule(lr){6-10} \cmidrule(lr){11-15}
\makecell[l]{Sequence} & V & Overrep & High & Low &
\makecell[l]{Sequence} & V & Overrep & With & Without &
\makecell[l]{Sequence} & V & Overrep & AU & MY \\
\midrule
\texttt{ST-RRRI-RQ}        & 0.132 & High & 39.0\% & 25.6\% &
\texttt{CP-ST-RQ-ST}       & 0.111 & With  & 35.6\% & 25.2\% &
\texttt{SS-RQ-RQ-PI}       & 0.150 & MY   & 38.9\% & 54.3\% \\
\texttt{ST-SS-RRRI}        & 0.130 & High & 37.0\% & 24.0\% &
\texttt{CP-RQ-CC-ST}       & 0.106 & With  & 32.4\% & 22.8\% &
\texttt{CP-CP-RQ-PI}       & 0.146 & MY   & 38.5\% & 53.4\% \\
\texttt{CC-ST-RRRI}        & 0.130 & High & 40.7\% & 27.4\% &
\texttt{CP-RQ-CC-PI}       & 0.099 & With  & 34.3\% & 25.1\% &
\texttt{SS-CP-RQ-PI}       & 0.143 & MY   & 38.8\% & 53.4\% \\
\texttt{CC-SS-RRRI}        & 0.128 & High & 44.8\% & 31.3\% &
\texttt{RRRI-ST-RQ-CC}     & 0.097 & With  & 26.8\% & 18.5\% &
\texttt{SS-RQ-PI}          & 0.141 & MY   & 40.8\% & 55.3\% \\
\texttt{ST-RRRI-SS}        & 0.127 & High & 32.5\% & 20.2\% &
\texttt{ST-CP-RQ-ST}       & 0.097 & With  & 38.6\% & 29.2\% &
\texttt{RQ-SS-RQ-PI}       & 0.141 & MY   & 40.8\% & 55.3\% \\
\texttt{ST-SS-RRRI-SS}     & 0.126 & High & 27.8\% & 16.3\% &
\texttt{CP-CP-RRRI-ST}     & 0.097 & With  & 27.8\% & 19.4\% &
\texttt{CP-RQ-RQ-PI}       & 0.140 & MY   & 38.9\% & 53.3\% \\
\texttt{CC-RQ-SS-RRRI}     & 0.124 & High & 43.9\% & 31.0\% &
\texttt{RRRI-ST-CC}        & 0.096 & With  & 30.8\% & 22.2\% &
\texttt{SS-SS-RQ-PI}       & 0.140 & MY   & 39.2\% & 53.6\% \\
\texttt{SS-RRRI-RQ}        & 0.124 & High & 42.6\% & 29.8\% &
\texttt{RQ-CC-CC-PI}       & 0.096 & With  & 39.6\% & 30.3\% &
\texttt{CP-SS-RQ-PI}       & 0.140 & MY   & 38.5\% & 52.8\% \\
\texttt{CC-SS-RRRI-SS}     & 0.122 & High & 34.9\% & 22.8\% &
\texttt{CC-CP-RQ-ST}       & 0.095 & With  & 45.8\% & 36.3\% &
\texttt{RQ-CP-RQ-PI}       & 0.139 & MY   & 40.0\% & 54.3\% \\
\texttt{RQ-RQ-SS-RRRI}     & 0.121 & High & 45.7\% & 32.9\% &
\texttt{RQ-RRRI-PI}        & 0.095 & With  & 32.5\% & 23.9\% &
\texttt{CP-RQ-PI}          & 0.138 & MY   & 40.1\% & 54.3\% \\
\bottomrule
\end{tabular}
\end{threeparttable}
\end{table*}

\section{Results}
\subsection {Hypotheses Related to Performance and Clinical Reasoning (H1.1-H1.3)}
Analysis of inquiry code frequencies (Table~\ref{tab:code_dist_overview}) revealed key differences between performance groups. High performers showed significantly higher occurrence frequencies of \texttt{RRRI} (Recognising and Responding to Relevant Information; 3.5\% vs 2.6\%), \texttt{ST} (Statement; 8.5\% vs 7.3\%), and \texttt{CC} (Chitchat; 15.5\% vs 14.7\%), while low performers demonstrated significantly higher frequencies of \texttt{RQ} (Routine Question; 38.2\% vs 36.9\%) and \texttt{CP} (Checking with Patient; 21.1\% vs 19.9\%) (all $p<0.05$). These frequency differences provided initial evidence for hypotheses H1.1-1.3, with higher information recognition and social elements in high performers versus greater reliance on routine questioning in low performers. Building on these frequency patterns, along the X-axis (MR1) of the comparative ENA plot (Figure~\ref{fig:ena} (a)), the high-performer group (Mdn=0.04, $N$=223) was significantly different from the low-performer group (Mdn=-0.06, $N$=100, $U$=8096.00, $p$<0.001, $r$=0.27). No significant difference was observed along the Y-axis (SVD2, $p$=0.65). The subtraction plot revealed that high performers' connections centered around \texttt{RRRI}, with particularly strong links to \texttt{CC}, \texttt{ST}, and \texttt{SS} (Specifying Symptoms) (red). This aligned with SPM results (Table~\ref{tab:h1h4h5_sm}): all the top ten significant inquiry sequences were overrepresented in the high-performer group. These discriminant inquiry sequences (Cramér's V=0.121-0.132) consistently featured \texttt{RRRI} as a central element, including \texttt{ST-RRRI-RQ}, \texttt{SS-RRRI-RQ}, and \texttt{ST-SS-RRRI-SS}. This pattern reflects the dense \texttt{RRRI}-centered connections in the ENA plot. It indicates integration of information recognition (significantly higher frequency) with summarization and symptom exploration and thus providing strong evidence to support H1.2. Similarly, \texttt{CC-SS-RRRI} and \texttt{CC-RQ-SS-RRRI} demonstrated how rapport-building provides foundation for clinical inquiry, reflected in the strong \texttt{CC} node connections in the ENA network and thus supporting H1.3. In contrast, low performers showed no significant sequences in the SPM analysis and their ENA networks remained confined to the procedural triangle of \texttt{RQ}, \texttt{SS}, and \texttt{CP} (blue lines in Figure \ref{fig:ena} (a)), indicating their questioning remained at the routine level without progression. Together, these results supported all three hypotheses (H1.1-H1.3).

\subsection{Hypotheses Related to First-Language Background and Clinical Reasoning (H2.1-H2.2)}
Analysis of inquiry code frequencies (Table~\ref{tab:code_dist_overview}) reveals distinct patterns between L1 and EAL speakers. L1 speakers showed significantly higher frequencies of patient-oriented codes---those reflecting rapport-building, collaborative dialogue, and responsive engagement: \texttt{CC} (Chitchat; 16.2\% vs 14.8\%), \texttt{ST} (Statement; 8.7\% vs 7.9\%), and \texttt{RRRI} (Recognising and Responding to Relevant Information; 3.5\% vs 3.0\%), while EAL learners demonstrated significantly higher \texttt{RQ} (Routine Question; 38.2\% vs 35.5\%) frequency (all $p<0.05$). Notably, \texttt{CP} (Checking with Patient) showed no significant difference between groups (20.5\% vs 20.2\%), suggesting verification behaviors alone do not fully characterize EAL learners' patterns. These frequency differences provide initial evidence, particularly supporting H2.2 through higher social elements in L1 speakers, while H2.1 requires further examination. Along the X-axis (MR1) of the comparative ENA plot (Figure~\ref{fig:ena} (b)), L1 speakers (Mdn=0.07, $N$=116) was significantly different from EAL learners (Mdn=-0.03, $N$=207, $U$=8706.00, $p$<0.001, $r$=0.27). No significant difference was observed along the Y-axis (SVD2, $p$=0.89). The subtraction plot revealed distinct co-occurrence patterns: L1 speakers showed strong \texttt{CC}-\texttt{RQ} and \texttt{CC}-\texttt{SS} connections, with \texttt{CC} also extending to \texttt{ST} and \texttt{CP} (red). This shows integration of social rapport (significantly higher \texttt{CC} frequency) across multiple inquiry types (supporting H2.2). In contrast, EAL learners exhibited prominent \texttt{CP}-\texttt{RQ}/\texttt{RQ}-\texttt{SS}/\texttt{SS}-\texttt{CP} connections (blue). Despite similar \texttt{CP} frequencies across groups, these connections reveal that EAL learners rely on verification to connect routine questions with symptom exploration, forming repetitive cycles rather than progressive inquiry and thus providing certain evidence for H2.1.

However, SPM did not identify any significant discriminant sequences between the first-language groups. This absence of sequential patterns, combined with the frequency and network differences, suggests that language background shapes inquiry through pairwise connections rather than extended temporal sequences. These results together provide partial support for both hypotheses: L1 speakers demonstrated more patient-oriented elements through higher social rapport frequencies and integrated network connections (H2.2), while EAL learners' verification-focused patterns appear only in network connections (H2.1).

\subsection{Hypotheses Related to Prior Pharmacy Work Experience and Clinical Reasoning (H3)}
Analysis of inquiry code frequencies (Table~\ref{tab:code_dist_overview}) revealed distinct patterns between students with and without prior pharmacy experience. Experienced students showed significantly higher frequencies of \texttt{CC} (Chitchat; 15.9\% vs 14.6\%), \texttt{ST} (Statement; 8.5\% vs 7.7\%), and \texttt{RRRI} (Recognising and Responding to Relevant Information; 3.4\% vs 3.0\%), while inexperienced students demonstrated significantly higher frequencies of \texttt{RQ} (Routine Question; 37.8\% vs 36.9\%) and \texttt{SS} (Specifying Symptoms; 14.5\% vs 12.8\%) (all $p<0.05$). These patterns suggest experienced students engaged more in social rapport and information synthesis, aligning with H3.1 and H3.2, respectively. Building on this, along the X axis (MR1) of the comparative ENA plot (Figure~\ref{fig:ena} (c)), students with prior pharmacy experience (Mdn=0.03, $N$=177) were significantly different from those without (Mdn=-0.01, $N$=146, $U$=10454.00, $p$<0.001, $r$=0.19). No significant difference was observed along the Y-axis (SVD2, $p$ = 0.67). The subtraction plot revealed experienced students' connections centered around \texttt{ST}, \texttt{RRRI}, and \texttt{CC} (red), with \texttt{ST} serving as a hub connecting to multiple inquiry types. This pattern aligned with SPM results: the ten significant sequences were all overrepresented in the experienced group (Table~\ref{tab:h1h4h5_sm}, Cramér's V=0.095-0.111). The patterns revealed \texttt{ST} as a structuring device: \texttt{CP-ST-RQ-ST} and \texttt{ST-CP-RQ-ST} showed how \texttt{ST} brackets surrounding questions to organize the flow of inquiry, consistent with \texttt{ST}'s prominent position in the ENA network (red) (supporting H3.2). Additionally, patterns like \texttt{CC-CP-RQ-ST} and \texttt{CP-RQ-CC-PI} demonstrated how experienced students integrated social elements throughout their clinical inquiries. This aligned with the \texttt{CC} connections to multiple codes (\texttt{RRRI}, \texttt{CP}, \texttt{PI}, \texttt{RQ}) visible in the ENA network (supporting H3.1). In contrast, no significant sequences were overrepresented in the inexperienced group, whose ENA networks remained confined to strong \texttt{CP}-\texttt{SS} and weaker \texttt{SS}-\texttt{RQ} connections. Together, these results supported both hypotheses (H3.1-H3.2).

\subsection{Hypotheses Related to Institutional Context and Clinical Reasoning (H4)}
Analysis of inquiry code frequencies (Table~\ref{tab:code_dist_overview}) revealed distinct institutional patterns. AU students showed significantly higher frequencies of \texttt{RRRI} (Recognising and Responding to Relevant Information; 3.4\% vs 2.9\%), \texttt{ST} (Statement; 8.6\% vs 7.5\%), \texttt{CP} (Checking with Patient; 20.8\% vs 19.6\%), and \texttt{CC} (Chitchat; 15.7\% vs 14.6\%), while MY students demonstrated significantly higher frequencies of \texttt{RQ} (Routine Question; 39.0\% vs 36.1\%), \texttt{SS} (Specifying Symptoms; 14.0\% vs 13.3\%), and \texttt{PI} (Professional Instructions; 2.4\% vs 2.1\%) (all $p<0.05$). These proportions suggest AU students emphasized social-clinical integration and information synthesis, while MY students focused more on structured procedural inquiry with professional instructions. Building on this, along the X-axis (MR1) of the comparative ENA plot (Figure~\ref{fig:ena} (d)), the AU students (Mdn=0.02, $N$=211) was significantly different from the MY students (Mdn=-0.07, $N$=112, $U$=8159.00, $p$<0.001, $r$=0.31). No significant difference was observed along the Y-axis (SVD2, $p$=0.77). The subtraction plot revealed MY students' connections forming a procedural triangle among \texttt{RQ}, \texttt{SS}, and \texttt{CP} with extensions to \texttt{PI} (blue), consistent with their significantly higher frequencies in \texttt{RQ}, \texttt{SS}, and \texttt{PI}. This pattern aligned with SPM results: the ten significant sequences were all overrepresented in the MY group (Table~\ref{tab:h1h4h5_sm}, Cramér's V=0.138-0.150). Notably, all sequences converged on \texttt{PI} as the endpoint. For instance, \texttt{SS-RQ-RQ-PI}, \texttt{CP-CP-RQ-PI}, and \texttt{SS-SS-RQ-PI} illustrated a gradual build-up from repeated routine and procedural questions toward professional instructions, reflecting the strong \texttt{PI} connections in the ENA network. In contrast, no significant sequences were overrepresented in the AU group, whose inquiry patterns are better characterized by the ENA's \texttt{CC}-\texttt{ST} and \texttt{CC}-\texttt{CP} connections, with \texttt{RRRI} serving as a relatively small hub linking social and structural elements. This network structure was consistent with AU students' higher frequencies in \texttt{CC}, \texttt{ST}, \texttt{CP}, and \texttt{RRRI}. It suggested transitions between rapport-building and clinical inquiry, integrating information synthesis, verification, and recognition. These results supported H4: students from different institutional contexts produced distinct inquiry patterns, with MY students employing procedural progressions toward professional instructions while AU students emphasized social-clinical integration through pairwise connections.

\section{Discussion and Conclusion}
\subsection{Interpretation of Results}
The exploration of inquiry patterns in GenAI-supported medication history-taking has revealed how students develop clinical communication competencies through unrestricted natural language interactions. Although effect sizes for sequences were relatively small (Cramér's V ranging from 0.095 to 0.150 across hypotheses), prior research has emphasized that frequently recurring patterns may nonetheless capture meaningful behavioral strategies \citep{gasevic2017detecting}. This observation is particularly relevant given that all the hypotheses revealed statistically significant differences despite modest effect sizes, indicating distinct inquiry patterns across student groups characterized by their performance levels and demographic attributes.

Our findings largely align with existing research on clinical communication education: high performers demonstrated more transitions between diverse inquiry types and centered their patterns around information recognition \citep{furstenberg2020assessing}, experienced students employed metacognitive scaffolding through statement markers \citep{higgs2024clinical}, and institutional differences reflected broader healthcare contexts \citep{jessee2016influences}. However, several distinctive patterns emerged that warrant deeper discussion.

First, a striking finding was the complete absence of significant SPM patterns in both the low-performer group and the inexperienced student group (those without prior pharmacy work experience). This suggests not that they adopted problematic inquiry patterns, but rather that they lacked strategic awareness of which patterns could facilitate improvement. In conventional training settings, instructors typically provide scaffolding that helps even low performers develop recognizable patterns \citep{van2010scaffolding}. By contrast, GenAI systems do not naturally offer such pedagogical guidance. In these open-ended environments, students must autonomously recognize and deploy effective inquiry strategies; without this capacity, they become disoriented in their questioning approach \citep{pozdniakov2022question}. Importantly, this does not mean GenAI worsens performance. Rather, it exposes the absence of strategic awareness that might remain hidden in traditional training contexts where instructor guidance masks these deficiencies---in traditional pharmacy education, instructors commonly observe student interactions and provide immediate feedback through modeling, strategic questioning, or direct intervention when learners struggle  \citep{smith2022developing}. This highlights how strategic awareness and content knowledge are equally essential in open learning environments \citep{li2024analytics}, suggesting that GenAI-based training requires explicit strategy instruction rather than simply providing unrestricted ``free practice'' opportunities.

Second, our complementary learning analytics methods revealed insights invisible to frequency analysis alone. While CP (Checking with Patient) frequencies showed no significant difference between language groups (20.5\% vs 20.2\%), ENA revealed its central position in EAL learners' networks, connecting routine questions with symptom exploration. This demonstrates how learning analytics can uncover meaningful patterns beyond simple frequency counts. EAL learners used verification not more frequently but differently, as a bridging strategy between inquiry types. Similarly, while RRRI represented only 3.5\% of high performers' dialogue acts (compared to 2.6\% in low performers), it served as a primary hub connecting diverse inquiry elements in the ENA network. Despite this relatively small proportion among all inquiry types, RRRI-based sequences were exclusively overrepresented in high performers. This suggests that the strategic positioning of information recognition within inquiry patterns is as important as its frequency of occurrence, indicating that strategic deployment and raw frequency work together in distinguishing successful medication history-taking practice.

Third, the partial support for language-related hypotheses reveals GenAI's complex role in addressing linguistic barriers. While cognitive load theory predicts that EAL learners would show restricted patterns due to linguistic processing demands \citep{venkat2020using}, our findings suggest a nuanced picture. Differences between EAL learners and L1 speakers appeared only in pairwise connections---EAL learners showed verification cycles through CP-RQ-SS connections, while L1 speakers integrated social rapport (CC) across multiple inquiry types. The absence of extended sequential patterns in SPM analysis for both groups suggests the GenAI environment may have moderated these linguistic effects. Unlike supervised learning environments that can negatively impact less confident students' performance \citep{paradis2017beyond}, GenAI-supported practice provides consistent, clear responses in a low-anxiety setting. This indicates that while cognitive load still shapes basic co-occurrence strategies (pairwise connections), GenAI environments may partially compensate for linguistic challenges that would otherwise extend to more complex sequential patterns.

Finally, the overlap between high performers and experienced students, with both of these two groups showing RRRI as central hubs and demonstrating structured inquiry patterns, suggests that prior experience may inherently contribute to better performance outcomes. This aligns with established evidence that clinical experience enhances performance in traditional settings \citep{choi2023prior}.

\subsection{Implications}
The inquiry patterns identified in this study provide educators with a new lens for understanding medication history-taking training through learning analytics rather than traditional observation alone. Our combined use of ENA and SPM demonstrates how these methods can reveal patterns invisible to traditional assessment approaches. A particularly notable finding was the complete absence of strategic patterns in low performers, suggesting they lacked awareness of which patterns could facilitate improvement rather than simply adopting ineffective approaches. Future GenAI-supported platforms could leverage these behavioral patterns to help students move beyond knowing what questions to ask, guiding them toward strategically connecting different inquiry types and developing effective sequential progressions. Beyond the specific context of pharmacy education, this study contributes to learning analytics methodology by demonstrating how combining ENA and SPM can reveal distinct behavioral patterns in unstructured GenAI-mediated dialogues---a growing need as health professions education increasingly adopts GenAI-powered training platforms for clinical skill development.

Our findings indicate that coded dialogue traces from GenAI platforms could enable real-time analytics dashboards that automatically detect unproductive patterns and trigger just-in-time interventions, similar to how multimodal analytics systems close the feedback loop for end-users in collaborative learning contexts \citep{zhao2024epistemic}. For instance, when the system detects a student has missed an opportunity to respond to clinically significant information from the VP, a mentor pharmacist agent could interject: \textit{``Wait. The patient just mentioned they stopped taking their blood pressure medication. This is important information you should explore. You could ask why they stopped, when this happened, and whether they experienced any side effects.''} This establishes a learning analytics foundation for formative assessment in GenAI-supported clinical training, providing actionable interventions as students' inquiry patterns unfold \citep{zhao2024epistemic, pozdniakov2022question}. Based on these automated interventions, educators can leverage the dashboard to zoom in and track how individual students apply different inquiry types across multiple practice sessions, such as observing whether students remain stuck in confined procedural inquiries or successfully integrate social rapport throughout their clinical questioning. At the class level, when educators identify common weaknesses displayed by the majority of students, they can incorporate these insights to enhance their teaching practices. Through the integration of ENA and SPM methods in the analytics dashboard, students can also compare their dialogue patterns against exemplar performances by examining visualizations that pinpoint specific moments where different inquiry strategies should have been applied. These fine-grained analytical capabilities would not have been feasible without learning analytics, empowering both educators and students to enhance clinical communication skills development.

The personalized nature of GenAI environments particularly benefits students with diverse backgrounds. For EAL learners, these platforms provide low-anxiety practice spaces that are particularly valuable at the beginning stages of clinical communication skill development, allowing them to build foundational competencies before they must face the linguistic pressures inherent in real patient interactions with diverse language backgrounds. Educators can encourage these students to practice integrating social rapport elements into their inquiries, which is a skill our findings showed they often lacked compared to L1 speakers. To address the institutional variations identified, future GenAI systems should incorporate diverse VP avatars that reflect varied communication expectations and clinical interaction styles, consistent with evidence that network-analytic approaches reveal systematic differences in learners' interaction structures across groups \citep{yan2023sena}, allowing all students to develop adaptive communication strategies aligned with their respective institutional contexts and healthcare practices. 

\subsection{Limitations and Future Directions}
Although this study, through learning analytics, has revealed how prior research findings translate to GenAI-assisted clinical training contexts, several limitations should be noted. First, our sample was drawn from a single university's pharmacy program (albeit across two campuses), limiting generalizability to other institutions or healthcare disciplines. Second, our analysis relied on seven inquiry codes, which may not fully capture the nuanced interactional competencies in clinical communication. Future work could develop more fine-grained coding schemes drawing on frameworks for professional communication competencies \citep{dai2024interactional}. Third, while our automated coding model based on LLaMA achieved an overall accuracy of 0.800 and weighted F1-score of 0.805, prediction accuracies varied across categories, with some codes showing relatively lower performance. Future improvements in training data and methods could enhance model performance. Finally, the cross-sectional study design prevents causal inferences about whether specific patterns lead to better outcomes or whether high performers naturally adopt these patterns. Future research should employ longitudinal designs with multi-institutional samples and compare these AI-assisted practices with traditional supervised approaches to better understand how such environments shape clinical reasoning development.

\bibliographystyle{ACM-Reference-Format}
\bibliography{LAK}

\end{document}